# Cross-Media Usage of Social Big Data for Emergency Services and Volunteer Communities

Approaches, Development and Challenges of Multi-Platform Social Media Services

Marc-André Kaufhold[1,2]
[1]Institute for Information Systems,
Computer Supported Cooperative
Work and Social Media (CSCW)
University of Siegen
Siegen, NRW, Germany
marc.kaufhold@uni-siegen.de

Christian Reuter[2]
[2]Science and Technology for
Peace and Security (PEASEC)
Technische Universität Darmstadt
Darmstadt, Hesse, Germany
reuter@peasec.tu-darmstadt.de

Thomas Ludwig[3]
[3]Institute for Information Systems,
Cyber-Physical Systems (CPS)
University of Siegen
Siegen, NRW, Germany
thomas.ludwig@uni-siegen.de

**ABSTRACT**

The use of social media is ubiquitous and nowadays well-established in our everyday life, but increasingly also before, during or after emergencies. The produced data is spread across several types of social media and can be used by different actors, such as emergency services or volunteer communities. There are already systems available that support the process of gathering, analysing and distributing information through social media. However, dependent on the goal of analysis, the analysis methods and available systems are limited based on technical or business-oriented restrictions. This paper presents the design of a cross-platform Social Media API, which was integrated and evaluated within multiple emergency scenarios. Based on the lessons learned, we outline the core challenges from the practical development and theoretical findings, focusing (1) cross-platform gathering and data management, (2) trustability and information quality, (3) tailorability and adjustable data operations, and (4) queries, performance, and technical development.

## 1 Introduction

Social media are nowadays widely established for everyday life uses, such as self-promotion, relationship building, news posting or information searching [1], but also as a source for information during natural and man-made crises and conflicts [2]–[5]. Facebook played a significant role in the "Arabic Spring" uprising, in which the tool facilitated the communication and interaction between participants of political protests [6], [7]. Twitter served as an important companion during natural disasters such as Hurricane Sandy in 2012, where it was used to communicate as well as to seek and offer emergency information [8]. Furthermore, YouTube provided a platform for more than 1,800 campaign-related videos of the later US President Barack Obama in 2008 that gained more than 50 million views [9].

Social media services coupled with smartphone technologies offer a permanent opportunity to create and gather information anywhere at any time in large quantities. Although the emerging 'social big data' [10] potentially boosts situational awareness for emergency services [11] and allows bidirectional communication with affected citizens as well as volunteer communities [12], [13], 'social media analytics' is still challenging [14]. Interesting data is spread across several different social media services [15], [16], the access to the data is often limited due to technical or business-oriented restrictions [17], issues of chaotic use, information overload and quality arise [18]–[20], and the analysis must adapt to the kind of data exchange format [21].

Within our article, we review related work of cross-platform social media data gathering and processing with regard to different purposes and provide an overview of existing technologies. Although we motivate the relevance of social media analytics, the specific focus lies within challenges and technologies for gathering of social data. Therefore, we present a systematic comparison of existing social media gathering platforms to highlight their contributions and limitations (section 2). Based on their shortcomings, we then introduce the architecture of a 'Social Media API', which has been designed and implemented during our three-years European project EmerGent, in terms of used specifications and technology (section 3). It allows to gather, process, store and re-query social data from Facebook, Twitter, Google+, Instagram, and YouTube.

To shed light on the practical applicability, potentials, and obstacles of the Social Media API, we deployed the interface in different applications over three years in the field of crisis management (section 4). We outline our experiences and discuss the results regarding the cross-platform gathering of social media data and their preparation for further analysis (section 5). The discussion entails lessons learned about mapping cross-platform social data into a uniform structure, technical limitations of the internal architecture and external social data provider API's, as well as contextual limitations (section 6).

## 2 Background and Related Work

In this section, we review related work of cross-platform social media data gathering, processing, and analysis as well as requirements for and a comparison of existing social media gathering services.

### 2.1 Social Media Analytics

Due to the increasing dissemination of mobile devices and the growing use of social media, global data traffic has steadily increased in the last years and will grow even faster until 2020 [22]. Application programming interfaces (APIs) make it possible to retrieve and process large quantities of data. This data is often summarized as *big social data* [23] which "is any high-volume, high-velocity, high-variety and/or highly semantic data that is generated from technology-mediated social interactions and actions in digital realm, and which can be collected and analysed to model social interactions and behaviour" [10]. Since the data are highly heterogeneous and interdependent, *social media analytics* is applied, which describes the process of gathering, analysing, and interpreting social media data [24]. The focus is on generating new knowledge of actors, entities, and social media relations to support both decision-making and execution of activities. Not only social media such as Facebook, Google+ and Twitter, but also blogs, forums and other platforms characterized by the nature and extent of user-generated content serve as an information source [25]. Furthermore, social media analytics is a holistic approach as not only analytics functions, but also monitoring is part of it. In this context, the focus is on quantifying selected indicators to capture arising trends and moods on social media [26]. Thus, methods enabling an aggregated view on information assets must be applied, for instance, by utilizing the potentials of data mining and machine learning [27], [28].

There are several characteristics of social big data that have to be taken into account: Firstly, social media comprise a significant number of users, relationships, and user-generated content such as posts and comments [29]. Therefore, vast computing and storage capacity is necessary [26]. Secondly, misspelling, shortcuts and ASCII emoticons impede the work of mining software [29]. The lack of structure can either be corrected automatically, which is defined as data cleaning, cleansing or scrubbing, or it can be used as a basis for quality analyses [26]. Thirdly, symbols, ambiguous expressions, irony and many more aspects depend on context [27], but metadata of users is often non-existent, and posts are linked since the creators are socially connected [29]. Conventional data mining does not regard this aspect. Fourthly, there are diverse access methods to social platforms such as different social media APIs [27] with various technical and business model oriented restrictions [17] and foreign languages and expressions that challenge the access [26]. Finally, ethical consequences must be considered: What consequences arise from collecting, processing, using and reporting data, even if the data principally is "public" [27]?

### 2.2 Data Gathering and Technical Limitations

When it comes to detecting relevant big social data, the retrieval of data from social media platforms is the first step, whereupon data can be pre-processed and analysed. To apply analytical methods on a wider data basis makes it necessary also to consider both the underlying data structure and its sources. Social media platforms regulate the access to information from their systems via APIs. Since the access to these platforms is controlled highly heterogeneously, different quantitative and temporal rate limitations for the query of large data amounts arise [17]. For instance, groups and pages on Facebook are obstructive as they usually restrict access to those being member or follower of the group or page. Therefore, access limitations are a real problem for pre-processing in the analysis process as long as temporal dependencies play a very significant role. Palen and Vieweg [30] illustrate the necessity of a real-time-based data collection for valid evaluation of crisis-related events.

Consequently, heterogeneous and incomplete datasets may lead to a lower volume of useful data and a more complicated analysis. For example, most messages published on Twitter do not have an explicit position specification. Keywords of the messages can be linked with or without indicating the position to execute georeferenced clustering of all available tweets and to obtain the mean of the position data [31]. Extracting information using Natural Language Processing can be a further possibility to supplement or link incomplete datasets [32]. To request social media data, many social media monitoring providers combine crawl services with APIs [33]. If access to all data via API is desired, a significant cost factor emerges as social media platforms will not provide their data without monetizing it. However, this is not the only factor determining the costs [26]. On the one hand, the software for the acquisition and evaluation of data must either be developed or purchased. On the other hand, not only sufficient computing and storage capacity but also warranted big data security is needed.

### 2.3 Existing Services for Social Media Gathering

Not only freely accessible sources such as Google Trends, but also commercial sources enable the required data to be accessed [26]. For example, *OneAll* provides an API which not only unites more than 30 social media services but also consolidates the most powerful social functions in one solution. While the basic functionalities are free, there are different packages for higher demands that are more comprehensive and require payment of a fee. However, only those social media posts can be accessed via the API that have also been published with *OneAll*. Similarly, *Social Mention* is a search and analysis platform which aggregates and provides user-generated content from more than 100 different social networks based on a keyword search. However, relevant search parameters that are required for a sufficient overview of social media are not provided. Firstly, the search interval can be specified only very vaguely and secondly, the system does not support any coordinate-based geo search.

*GNIP* is Twitter's enterprise API platform and provides several products delivering social data to businesses including

PowerTrack, a filtering API, and Decahose, that produces a 10% sample of Tweets in real time. GNIP provides access to a wide range of popular social media services such as Tumblr, Foursquare, WordPress, and Disqus. *DataSift* is a UK-based company that enables companies to integrate social, blog and news data in a single place. It captures, augments and delivers real-time and historical social media data from a variety of sources including Twitter, Facebook, Google+ and Instagram. It has a single API based on a querying language. Since 2015, DataSift maintains a partnership with Facebook. However, both DataSift and GNIP require an approval of social media services for the exact use case and usually it costs at least $3000/month to make this worthwhile. *BlueJay* is a service allowing real-time monitoring of tweets for law enforcement services. It has access to the full Twitter Firehose and a moderate subscription cost ($150/month).

Besides commercial solutions, research develops solutions for gathering and analysing social data. *EPIC Analyze*, established by the University of Colorado Boulder, is a comprehensive, scalable, and extensible data analysis environment for crisis informatics research [34]. They describe the goal of EPIC Analyse as "helping our analysts sift through enormous Twitter data sets to produce representative sets of tweets that they can use to answer socio-behavioural and sociolinguistic questions around mass emergency events". The underlying gathering component, EPIC Collect, focuses on the reliable collection of large amounts of Twitter data. Furthermore, the SocialSensor project developed a framework for enabling real-time multimedia indexing and search on the social web [35]. The publicly accessible *Stream Manager* monitors a set of seven social streams: Twitter, Facebook, Instagram, Google+, Flickr, Tumblr and YouTube to collect incoming content relevant to a keyword, a user or a location, using the corresponding API that is supported by each service. The Twitter API is implemented as a real-time service, whereas the other six act as polling consumers performing requests to the network in intervals. The framework also provides wrappers to MongoDB, Solr, and Lucene storages.

## 2.4 Requirements for and Comparison of Social Media Gathering Services

Based on the examination of existing services, we developed requirements for a cross-platform social media gathering service and present a comparison of existing platforms and tools that fulfil these requirements (partially) [36]. Dealing with the specifics of social media data, existing challenges, technical limitations and some requirements should be fulfilled by approaches for the gathering of social media:

- **Multi-Platform Support (MP)**: Due to the high variety of social media services that are used during emergencies, we cannot specify in advance, where the most relevant citizen-generated information will be. A requirement is therefore that a request allows access to multiple platforms.
- **Cross-Platform Usage (CP)**: The high variety of social media services requires a large variety of different accounts and information spaces. To reach most users, a requirement is to allow the gathering of citizen-generated information spread widely across social media services.
- **Data Superiority (DS)**: To analyse data, without any limitations on requests or queries, and treat them in an appropriated way, e.g. regarding national or organizational privacy and ethical issues, a requirement is that the flexibility of data storage on own servers must be guaranteed.
- **Crawl Service (CS)**: Besides single searches, a "crawl service" should allow gathering the data over a pre-defined period to continuously capture citizen-generated information in everyday life or during specific events, such as emergencies, in nearly real-time.
- **Location- and time-based data (LT)**: During events or emergencies, location- and time-based information are very important because they provide crucial contextual information. Since not all activities will contain location-based metadata, ways of extraction are to be considered.
- **Interoperability (IO)**: To process cross-platform social media data, the gathered data must be stored in a unified format. Moreover, the storage should follow an accepted specification to ensure interoperability of data, thus allowing third-party applications to integrate the data.

Comparing the analysis of existing tools with the raised requirements, there is a variety of cross-platform services to query social media data, but not all examined services fulfil the requirements. Table 1 depicts to what degree each service addresses the given requirements. Although Social Sensor fulfils most of the defined requirements, its implementation was made publicly available after the development of our own component already started. While services like OneAll and Social Mention have restrictions on data availability and filtering methods, the monthly fee for using GNIP or DataSift is too expensive for research projects, if service availability must be maintained over an extended period, and data is stored on external servers. Moreover, platforms like BlueJay and EPIC Analyse are restricted to Twitter. Therefore, implementing an own cross-platform service seemed to be the best solution for a) tailoring the artefacts, b) maintaining the functionality with internal expertise and c) enabling the best adaptability, extensibility, and interoperability for changing or enhanced usages.

|    | One All | Social Mention | GNIP | Data Sift | Blue Jay | EPIC Analyze | Social Sensor |
|----|---------|----------------|------|-----------|----------|--------------|---------------|
| MP | ✓       | ✓              | ✓    | ✓         | ✗        | ✗            | ✓             |
| CP | (✓)     | (✓)            | ✓    | ✓         | ✗        | ✗            | ✓             |
| DS | ✓       | ✓              | ✗    | ✗         | ✗        | (✓)          | ✓             |
| CS | ✗       | ✗              | ✓    | ✓         | (✓)      | ✓            | ✓             |
| LT | (✓)     | (✓)            | ✓    | (✓)       | ✓        | ✓            | ✓             |
| IO | ✗       | ✗              | ✗    | ✗         | ✗        | ✗            | ✗             |

**Table 1. Evaluation of Requirement Fulfilment**

## 3 Architecture of a Social Media API

The 'Social Media API' (SMA), which will be presented in this article the first time as a whole, allows to gather, process, store, and re-query social media data. Based on underlying social media, the SMA contains different services that are used by several client applications. Although it was developed as enabling technology for crisis management applications, its implementation allows to support a variety of use cases in different fields of application, e.g., to examine the impact of a product image within the field of market research. To enable access to social big data and allow subsequent analysis, our first step was to specify a service for gathering and processing social media content. With *gathering*, we refer to the ability to uniquely or continuously collect social media activities (e.g., messages, photos, videos) from different platforms (Facebook, Google+, Instagram, Twitter, and YouTube) in a unified manner using multiple searches or filter criteria, and *processing* means that the SMA is able to access, disseminate, enrich, manipulate, and store social media activities.

**Technology**. The SMA is realized as a service following the paradigm of a web-based, service-oriented architecture (SOA). It is a Java Tomcat application using the Jersey Framework for REST services and the MongoDB database via Hibernate Object/Grid Mapper (OGM) for document-oriented data management. The implementation allows using a local or remote instance of MongoDB. Several libraries facilitate the integration of social media platform APIs like Facebook Graph API or Twitter Search API. To overcome the diversity of data access and structures, all gathered social media entities are processed and stored according to the ActivityStreams 2.0 Core Syntax in JavaScript Object Notation (JSON).

| *Resource* | *[POST] /SocialMediaAPI/crawlService* |
|---|---|
| *Payload example (application/json)* | `{ "gathering": { "keyword": "berlin", "platforms": ["facebook","instagram","twitter","youtube"], "waitBetweenRequests": 10000 } }` |
| *Response examples (application/json)* | `{ "crawljobId": "ea8689d02f04b1b5b39e93d66adbb1ff580084" }` |

**Table 2. Initialization of a basic crawl job based on the parametrization of the payload**

**Endpoint Functionality**. SMA comprises two main services, each providing a multitude of service functions: The *Gathering Service* contains endpoints for gathering and loading social media activities. The main components are the Search service, enabling onetime search requests, and Crawl Service, which continuously queries new social media activities across a specified timeframe. Using the *Enrichment Service*, gathered social media activities are enriched with further computed and valuable metadata. For the initialization of a crawl job, a POST call is sent to the *crawlService* endpoint, which contains a payload matching the *application/json* Content-Type header. In its basic configuration, a keyword, at least one platform and an interval value to determine the timeframe between each gathering request are required; further configuration parameters, may be specified to filter the scope of the crawl job. To query gathered results of a certain crawl job, a GET call is sent to the *crawlService/{crawljobId}* endpoint with *{crawljobId}* being a concrete instance of an identifier. The identifier may be retrieved from the response of the crawl job's initialization or from the list of the *crawlService/allJobs* endpoint (Table 2).

**Data Specification**. The Activity Streams 2.0 Core Syntax (AS2) defines that "an activity is a semantic description of potential or completed actions" [37], which has at least a verb (type of activity, e.g. like, post, share), an actor (e.g. creator) and an object (e.g. an image or message object). There are already many verbs and object types defined within a specification, for instance, a place object may contain the attributes latitude, longitude, and altitude. Although the specification allows modelling the activities of liking, sharing and so on, there are no attributes designated to carry information like "20 users liked this post". While the specification may be extended with own verbs and object types, foreign implementations possibly have not enough knowledge to process them in an intended way. Activity objects must be encapsulated in a Collection object before returning them as a JSON object.

**Data Storage**. For storing and retrieving the collected data in and from a MongoDB database instance, we deploy the Java framework Hibernate OGM. We selected MongoDB as document-oriented NoSQL solution due to its good performance in reading, writing and deleting operations on large datasets and, compared to SQL solutions, flexible document schemas and the option of *sharding*, a method for distributing data across multiple instances or machines [38]. Furthermore, with the aid of the OGM tools (object-grid mapper), we could operate without direct database commands because they are encapsulated in the framework, e.g., as save, update or delete functions. The generation of the database scheme and the storage of the corresponding instances of the objects is done automatically. Only the annotations of the appropriate classes and their attributes are needed for Hibernate to transform the Java classes into database query commands. Moreover, we use a compound unique index based on the activity's platform and id to prevent duplicate activities on database level.

## 4 Fields of Deployment and Evaluations

In this section, we discuss the fields of deployment strongly connected to the presented client applications (section 4): The SMA serves as foundation for multiple analytical client applications. Although the individual applications [renamed for review] are already published in previous papers, this section concisely introduces each application and the core results of their evaluation to inform the upcoming analysis and discussion.

| Application | Evaluation Type | Part. |
|---|---|---|
| *Coordination Tool* [16]: XHELP | Scenario walkthrough, interviews | 20 |
| *Quality Tool* [39]: Social-QAS | Scenario walkthrough, interviews | 20 |
| *Monitoring Tool* [40]: CrowdMonitor | Scenario walkthrough, interviews | 28 |
| *Alert Tool* [19], [41]: Emergency Interface | Scenario walkthrough, interviews, functionality test | 33 |
| *Data Tool* [21]: Social Data Collector | Usability walkthrough, interviews | 12 |

**Table 3. Applications, Evaluation Types and Participants**

The evaluations (Table 3) were conducted in a comparable manner: During the exploration of these applications via cognitive or scenario-based walkthroughs, the participants were asked to 'think aloud' [42]. The semi-structured interviews which followed were intended to encourage reflection on the evaluation process. The results were audio-recorded and transcribed for further analysis. In our subsequent analysis, we employed "open" coding [43]. The philosophy behind these evaluations was derived from the notion of 'situated evaluation' [44] in which qualitative methods are used to draw conclusions about the real-world use of a technology using domain experts. However, although the conducted evaluations impacted the development of SMA, it must be noted that their focus was on the characteristics of the overlying applications.

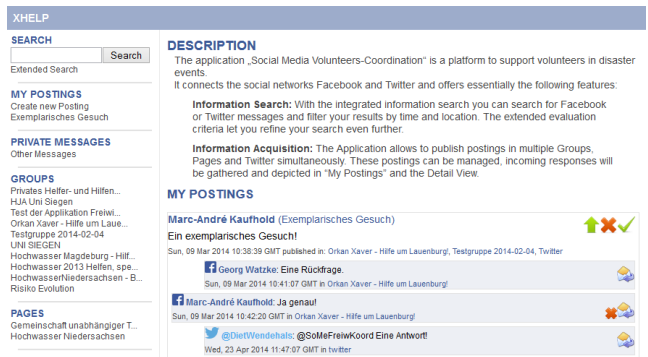

**Figure 1. Coordination Tool (CT): XHELP**

**Community Interaction and Dissemination of Social Media Messages**. XHELP is a Facebook application allowing users to acquire and distribute information across media (e.g., Facebook and Twitter) and channels (e.g., Facebook groups and pages) [16]. It provides an overview of their published posts, joined groups and liked pages. The central dashboard "My Postings" provides an overview of the user's communication threads, including the postings created with XHELP, but also postings and comments set up on the source platform. The users can (1) collapse or expand comments on a communication thread, (2) respond to any or delete own comments, and (3) remove or finalize own communication threads. Furthermore, the cross-media search allows to search for publicly and privately (e.g., if the user is a member of the respective Facebook group) disseminated Facebook and Twitter posts and filter them by time, geolocation and radius.

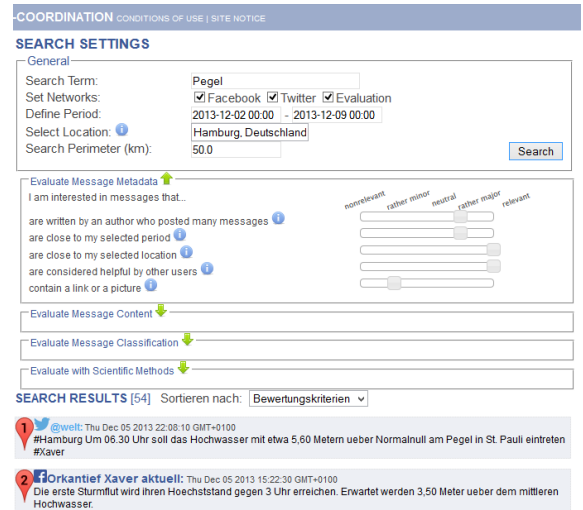

**Figure 2. Quality Tool (QT): Social-QAS as part of XHELP**

**Tailorable Data Filtering and Quality Assessment during Gathering**. Social-QAS aims at facilitating the assessment of social media content by the tailorable weighting of information quality criteria [39]. Because various circumstances call for different assessment methods, the possibility to combine these techniques could help to improve the quality assessment practice. The concept allows the assessment of content with 15 assessment methods, which are divided into four categories: metadata (e.g. follower, likes, media files); content (e.g. frequency of search keywords); message classification (e.g. sentiment analysis) and scientific methods (e.g. term frequency-inverse document frequency). If the end-user of an application based on Social-QAS has the possibility to choose and weight several assessment methods, a subjective quality of information can be determined. In an exemplary implementation into XHELP, it is possible to search for information by using different quality parameters to perform a quality assessment.

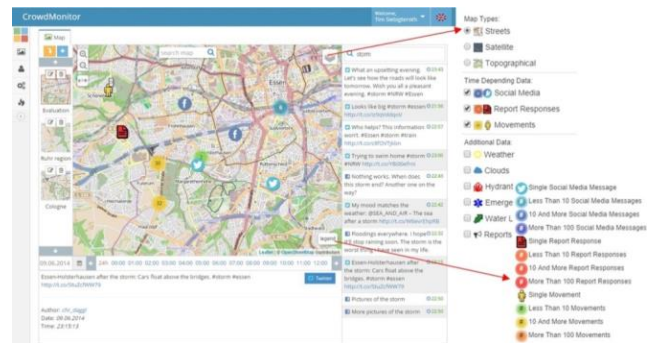

**Figure 3. Monitoring Tool (MT): CrowdMonitor**

**Combining Social Media Content and Civil On-Site Activities**. The web-application CrowdMonitor aims at enabling the sense-making of spontaneous volunteers' activities for emergency services during emergencies [40]. One of the challenges within crisis management is the awareness about activities of spontaneous volunteers and aligning these activities with those of official emergency services. To tackles this issue, it combines collective processes gathered via social media through the SMA with individual activities sensed with mobile devices. CrowdMonitor allows emergency services to passively collect and display social media information (from ordinary people without their knowledge). It further encompasses creating requests for particular information or targeted alerts, which can then be pushed to users of the mobile app (within a particular location). CrowdMonitor offers the potential of a synchronized view on mobile-gathered data and information collected through social media as well as subsequent possibilities for interacting with people to provide situation overviews.

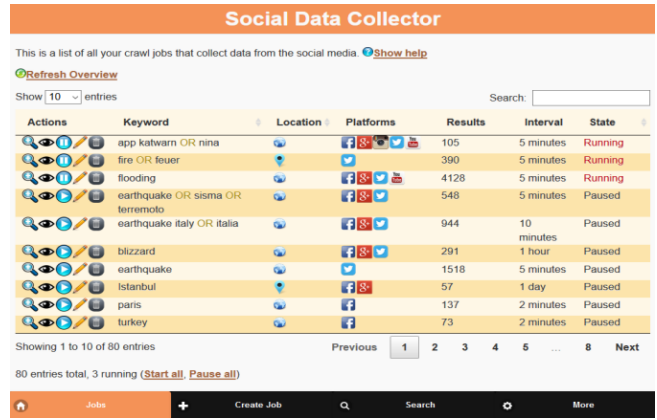

**Figure 5. Data Tool (DT): Social Data Collector**

**Creating Social Media Datasets for Analysis**. The Social Data Collector is a graphical user interface to create social media datasets for (crisis management) research [21]. It aims at supporting users in gathering and managing large amounts of data from different social media providers. This includes the day-to-day search for news as well as the continuous gathering and archiving over a longer period, whereby each search constitutes a separate collection. In terms of crisis management, a researcher might be interested in monitoring an actual emergency, but also certain events or locations where an emergency could or is likely to happen either to capture the outbreak of an emergency or, in conjunction with an analysis module, to gather and process indicators of an upcoming emergency. The Social Data Collector itself offers four main functionalities: a) An overview of all collections; b) a detailed view of the resulting posts of a collection; c) the initiation of interval-based search operations (crawl jobs); and d) the initiation of one-time search operations.

## 5 Discussion of Challenges

Based on the "lessons learned" we have made, we will present the potentials, but also obstacles of cross-platform gathering and analysing of social media data and their preparation for further analysis (Table 4). The discussion addresses the topics of mapping cross-platform social data into a uniform structure (specification level), technical limitations of the internal architecture and external social data provider API's (technology level) as well as contextual limitations regarding gathering the "why" of using social media services (event and user level).

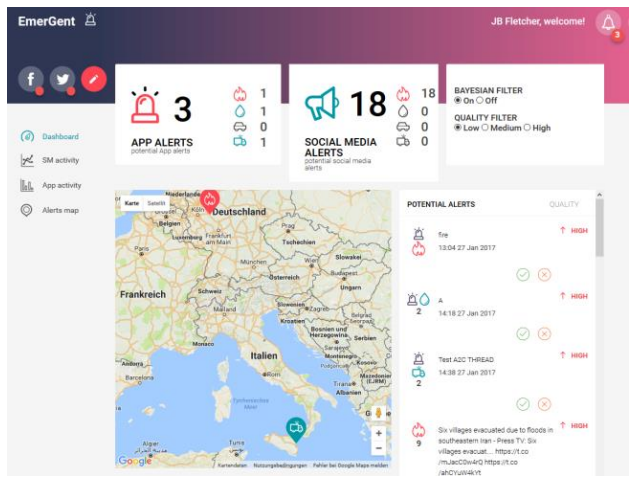

**Figure 4. Alert Tool (AT): Emergency Interface**

**Social Media Alerts for Emergency Services**. During an emergency, thousands of potentially relevant messages may come up, and this would lead to an information overload again. In the Emergency Interface, *alerts* aim at "transferring high volume, but unclear information content into low volume and rich content suitable for emergency services" [19], [41]. An alert is a set of classified messages sharing a similar context, which is of particular interest for emergency services. The context is defined by, but not limited to, attributes like date, time, location, full text, identified event types or language. Each alert consists of several messages from Twitter, Facebook, Google+, Instagram or YouTube if they belong to the same setting. Within the information mining process, relevant data is filtered, classified and automatically categorized. Furthermore, messages are evaluated for the information quality process to estimate different criteria like timeliness, understandability, believability or completeness. The user also has the possibility to adapt the process with specific keywords.

|    | Cross-Platform | Information Quality | Tailorability | Queries & Performance |
|----|----|----|----|----|
| CT | ✗ |   | ✗ |   |
| MT |   | ✗ |   |   |
| QT | ✗ | ✗ | ✗ |   |
| AT |   | ✗ |   | ✗ |
| DT | ✗ |   | ✗ | ✗ |

**Table 4. Key Evaluation Topics of Social Media API**

## 5.1 Cross-Platform Gathering and Management

The different social media platforms provide a multitude of information such as text, multimedia files, metadata (e.g. number of likes or retweets), and often they are handled in different ways (e.g. Twitter supports only 280 characters per message). The research effort aimed at finding a common and standard approach to represent and store information collected from social media, and we identified the Activity Streams 2.0 Core Syntax in conjunction with MongoDB as the best solution. On the one hand, the flexibility of the document-oriented approach allows storing distinct structured documents with different numbers of attributes. Using AS2, most attributes are stored according to an interoperable specification. On the other hand, regarding divergent metadata, the comparability and analysis of social media activities is restricted. Moreover, as not all attributes can be mapped to the AS2 specification, we needed to add a custom property mapping our special metadata (Table 5).

```
{
  "actor": {
    "content": "56, Ironie, eigene Meinung",
    "displayName": "anonymised",
    "id": "twitter:84430424271",
    "type": "person",
    "url": " https://goo.gl/QqV2q6"
  },
  "object": {
    "content": "RT @bzberlin: #Debüt mit 1:0 gegen @SERCWildWings https://t.co/UNlq698PlJ",
    "enrichedData": {
      "absFearFactor": 0,
      "absHappinessFactor": 0,
      "embeddedUrls": ["https://t.co/UNlq698PlJ"],
      "language": "de",
      "tags": ["Debüt"],
      "media": [{
        "mediaType": "image/jpeg",
        "type": "photo",
        "url": "https://goo.gl/QqV2q6"
      }],
      "mentions": ["bzberlin", "SERCWildWings"],
      "numOfCharacters": 133,
      "numOfWords": 11,
      "numRetweets": 3,
    },
    "id": "twitter:823724465664883940",
    "location": {
      "displayName": "Neunkirchen, Deutschland",
      "latitude": 50.78506988,
      "longitude": 8.00512706,
      "type": "place"
    },
    "startTime": "2017-02-01T10:30:47.000+01:00",
    "type": "post",
    "url": " https://goo.gl/QqV2q6"
  }
}
```

**Table 5. ActivityStreams with EnrichedData object**

Besides available data, there are two kinds of additionally valuable data: Firstly, some data is only available in certain social media but computable for others. For instance, embedded hyperlinks, mentions, or tags can be extracted from social media activities to get a comparable amount of data. Secondly, some required data regarding the assessment of quality is not available in any social media. The SMA, therefore, computes classification attributes (negative sentiment, positive sentiment, emoticon conversion, slang conversion), content attributes (number of characters, number of words, average length of words, words-to-sentences ratio, number of punctuation signs, number of syllables per word, entropy) and metadata attributes (hyperlinks, language, location, media files, tags) manually. However, standardization has its limits since it is an uneconomically endeavour trying to map each possible metadata across all social media into a single specification. Thus, a potential approach would be to store the native format, for instance, as a String attribute per activity, too. This would also help in case of delivering data to client applications that only support specific native formats.

## 5.2 Trustability and Information Quality

The development processes and the later evaluations of the applications show that issues of trustability and overall the information quality are always prevalent and occur on different levels. Questions arose whether the author of a social media post is trustful, whether the content of the message is relevant to a specific situation or whether the location of a posted message has an appropriate level of detail. What becomes evident is the context-dependent and highly individual character of information quality. The "fit of information to specific tasks is more important than generic assessments of information quality" [45]. It can be understood as "the quality information has for a certain purpose/goal", which depends heavily on personal reasons. Sometimes just data from the same place, data published after a specific point of time and messages of a specific type, such as pictures, are relevant. It differs not only from situation to situation but also according to the particular stage of an evolving situation. What one user needs to answer his/her questions can be irrelevant for other users – even in relation to the same or similar issues. It is therefore almost impossible to implement a pre-defined setting for information quality, and users must be able to dynamically tailor their data operations within a situation.

As one aspect of information quality, the trustability of social media data plays an important role. Analysing the content of social media messages is often confronted with the credibility of authors as well as (and as the current discussions about fake news show) the dissemination of rumours. Because inexperienced users often struggle with deciding whether a social media message is reliable, trustful and relevant, a cross-platform with the aim of gathering and analysing of social media data should allow a refined enrichment of social media data to get more contextual information. For example, such refined enhancement could be established through measurements of retweets/likes (e.g. Social-QAS). However, once the data gathered through cross social media platforms do not meet the context-dependent character of a situation or if information is not trustful (i.e. because an author is not trustful or the location is too broad), there have to be further ways of validating the (semi-)automatically processed information. Such validation could be possible by individualized reports from the scene, which extends the information base for situation assessment practices. By implementing pre-processed ranking and filtering of cross-gathered social media messages and enhanced options for further validation, the trustability and information quality can be increased. Nevertheless, one should consider that the more specific the search and the more accurate the filtering is, the less information one gets, which shrinks the information space.

## 5.3 Tailorability and Data Operations

Several evaluations showed that – to some extent – automation is necessary for dealing with social big data. However, some applications based on the SMA (especially Social-QAS and Social Data Collector) also indicated that a fully automated process could not meet the requirements. With the need for a kind of automatic provision and pre-processing of data and results, the need for tailoring arose. Consequently, SMA needs to be flexible enough by tailoring options for source platform selection and quality assessment criteria since situation assessment revealed itself to be very subjective. Subsequently, personal feelings, experience and the situation itself influence the information requirements. While gathering or analysing information and implementing information systems to encourage the task, there is always one important question: How can we realize information systems, which enable the automatic selection of relevant data and, simultaneously, grant end-users the option to adapt this automation, thus allowing tailorable quality assessment according to their requirements? This is even more essential if situations and the context of work vary and if practices develop over time. Concepts like Social-QAS will simplify the articulation of end-users concerning their needs.

| *Parameter* | *Type* | *Description* |
|---|---|---|
| keyword | String | Required. The search keywords. |
| platforms | String | Required. A list (Facebook, Google+, Instagram, Twitter, YouTube). |
| since/ until | Long | Search Service. Lower/upper bound of the searched timeframe (Unix time). |
| start/ end | String | Crawl Service. Starting/termination point of the crawl job (Unix time). |
| latitude/ longitude | Double | Latitude/longitude for geo search (decimal degree). |

**Table 6. Required and optional query parameters**

For SMA this leads to the requirements to provide an easy option to get results but also to give the user (with the help of different client applications) the opportunity to use all the metadata available to tailor the results needed for the situation. Of course, the designer can estimate a lot of 'usual' use cases; however, the practitioner should have the flexibility to regret these cases and to select data that is really needed in the particular situation. The use of assessment methods with multiple social media implies the need of a tailorable SMA. A key challenge is the provision of suitable service endpoints with sufficient filter parameters that behave consistently over heterogeneous social media. The flexibility of filtering depends on the providing APIs to some extent: While some social media APIs support location (Twitter, YouTube) and temporal (Facebook, Twitter, YouTube) filtering, it must be realized manually for the other ones. Nevertheless, given the quota limitations of social media, manual filtering always implies the previous gathering of results that do not match the filter criteria and is, therefore, less efficient than using native filter parameters.

After data is gathered and stored in the database, the access becomes an important factor to allow loading and post-processing of data. Given the job ID, social media activities of past crawl or search jobs can be loaded and filtered by count (amount of data returned) and offset (position of the first result to be returned) parameters. Alternatively, a list of activity IDs allows loading the desired social media activities explicitly. Yet, to enhance the tailorability of SMA to increase the flexibility for consuming client applications, the implementation of additional parameters is planned, e.g. keyword, platform, location and time-based filtering, or language. In this case, the efficiency and flexibility of filtering are dependent on the underlying database management solution.

## 5.4 Queries, Performance and Development

For querying data from multiple social media, the keyword parameter constitutes an issue since different social media API's process keywords differently and support various types and notations of logical query operators (e.g., AND, OR, NOT, phrases, parentheses). Here, the need for a unified query language and layer becomes apparent, which translates the unified query parameters into the platform-specific parameters. While Google+, Twitter, and YouTube use the same query syntax for the basic logical operators, Facebook's Graph API only supports AND conjunctions, and the Instagram API does not provide a keyword search, but only to search for single tags within the description of existing media. However, to enable further logical operators for Facebook and Instagram, they had to be implemented manually using the ANTLR, which is a parser generator for reading, processing, executing, or translating structured text or binary files [46]. For instance, the OR operator was translated into multiple API requests, and the NOT operator was realized by removing gathered messages containing the undesired keywords. This leads to a faster consumption of quota limits [17] due to the multiplication of performed requests (OR) and the gathering of undesired data (NOT). Consequently, especially with non-expensive approaches, it is possible to capture and process merely small portions of the high-volume social data.

Although the multiplication of requests also impacts the performance of the SMA, more severe performance issues became apparent during the technical evaluations of the Emergency Interface and Social Data Collector. During the 2015 Paris attacks and 2016 Brussels bombings, double-digits amounts of gigabytes or millions of social media activities were gathered and stored in our MongoDB. While the database size had no impact on inserting (writing) new documents, the performance during the query (reading) of arbitrary document collections was drained heavily. The subsequent investigations revealed that the combination of the regular index and the compound unique index, even without the working set, exceeded the size of the machine's Random-Access Memory (RAM). Thus, every time the database looked for a document whose identifier was not within the loaded index range, the RAM was rewritten completely with a new index range. Consequently, operations with a few seconds of execution time were extended to many minutes, which complies with the observations made by [34]. Here, solutions such as sharding (distributing data across multiple machines), more RAM or a more efficient indexing service, such as Redis, should be considered.

Moreover, the API's of social media providers are in development requiring continuous adaption to these changes. While some changes are small, e.g. the temporary requirement of a Google+ account to use YouTube, and are communicated to the developers early, some constitute a big impact on the implementation. For instance, an important issue has been raised due to the removal of Facebook's Public Post Search on April 30th, 2015, limiting the access to public data significantly. Furthermore, in June 2016, Instagram put all non-review apps into sandbox mode, which is limited to ten users with reduced API rate limits and whereby data is restricted to these ten users and the 20 most recent media from each of those users. To regain access to live data, the developers must perform a permission review, which requires submitting a video screencast with the login experience, the app to be in a production stage and to comply with one of three pre-defined uses cases. This poses a problem since the API does not provide a visible login experience, is in continuous development and does not comply with the three uses cases.

## 6 Conclusion

Social media is undoubted of high importance and raises several interests for diverse stakeholders within different application fields, such as emergency services and volunteer communities involved in emergencies [1], [2], [47]. Knowing the great importance of, and providing access to the big social data build the business model of different enterprises that offer the possibility to extract and use selected data in third party applications [10]. This article aims towards researching the opportunities for cross-media gathering and using of social big data to highlight approaches and limitations (that are partly known) and to present the development of a cross-platform API. We considered the existing approaches based on their eligibility within the dimensions multi- as well as cross-platforms, data superiority, the existence of crawl services, location- and time-based data as well as interoperability (Section 2). Although existing approaches and systems already addressed aspects of these dimensions, their detailed description of functionality behind the provision services didn't gain attention up till now.

Within the article, we presented a novel cross-platform Social Media API (SMA) that aims at supporting its users with functionalities covering the dimensions outlined above (Section 3). Although results of the individual case studies that are based on SMA have already been published [16], [19], [21], [39]–[41], the (a) API itself and the (b) discussion of challenges derived across all evaluations are novel and original contributions that have not been studied so far. The SMA supports multi-platform support, which is extensible with standardized interfaces and supports cross-platform gathering, processing and re-querying of acquired data. Additionally, it fulfils data superiority due to the possibility to deploy it on any application server. It also supports punctual and continuous searches as well as the capturing of meta-data. We deployed the SMA in various fields such as supporting volunteers across social media (XHELP), combining social media with movements of people for emergency services (CrowdMonitor), supporting tailorable quality assessment (Social-QAS), creating alerts for emergency services (Emergency Interface) and making the API usable for general emergency research (Social Data Collector) (Section 4).

| ID | Challenge | Description |
|---|---|---|
| C1 | Specification | Most, but not all metadata across social media can be stored according to an interoperable specification, such as ActivityStreams 2.0. |
| C2 | Comparability | The extraction or computation of metadata is sometimes required for comparing social data across platforms. |
| C3 | Interpretability | The interpretation upon objective metadata, such as trustability and information quality, is context-dependent and highly individual. |
| C4 | Tailorability | To meet end-users aims, the gathering of social data requires filter parameters, such as by location or time, which sometimes must be implemented manually. |
| C5 | Query Operators | Official social media APIs support different query operators, such as AND, OR and NOT, and the simulation of operators often is quota-heavy. |
| C6 | Performance | The export or query of large datasets from database requires intelligent and sophisticated storage solutions or performant systems. |
| C7 | External Change | Regular changes of official social media APIs, such as Facebook or Instagram APIs, require respective adaptions or can result in the loss of access to data. |

**Table 7. Summary of identified challenges**

Based on extensive evaluations, various discussion points appeared that influenced the (re-)design of the SMA, such as issues of (1) cross-platform gathering and data management, (2) trustability and information quality, (3) tailorability and adjustable data operations, and (4) queries, performance and technical development (Section 5). The challenges and requirements for a tool like the SMA are primarily defined by its client applications, research objectives and use cases.

However, we outline experienced challenges that occurred during the development and deployment for different use cases in Table 7. For example, in conjunction with the Emergency Interface, the SMA is part of a broader backend, which also includes components for data mining, information quality, information routing and alert generation [48]. Data provided from SMA was used to create and train an information quality model, which in the next step will be evaluated in comparison to real user input and, if necessary, adjusted accordingly. Due to the agile and regular refinement of client components and underlying models, the SMA undergoes a continuous process of change. Further potentials for emergency management lie within the implementation of community or event detection techniques, which would imply additional requirements [49], [50].

This, moreover, is influenced by regular adjustments of social media provider APIs, as described by the examples of Facebook as well as Instagram, and the changing landscape of existing social media providers. Although Twitter, for instance, provides easy data access with a reliable API, the access to Facebook data is limited. Research in the field of crisis informatics outlined that different social media are used for diverse purposes [15], which means the examination of various phenomena and (comparative) multi-platform analysis is affected due to the limitations of data access. Last, besides tracking the changing landscape of social media, different application domains may yield additional challenges or put emphasis on divergent core challenges.

## ACKNOWLEDGMENTS

This research was funded within the research project "EmerGent" of the European Union (FP7 No. 608352) and the research group "KontiKat" [51] of the German Federal Ministry of Education and Research (BMBF No. 13N14351)